\documentclass{eas}

\usepackage{graphicx}

\def\drt{d^3r}

\def\az{a_{0}}
\def\haz{\hat\az}

\def\l0{\ell_{0}}

\def\rar{\rightarrow}

\def\l{\lambda}
\def\f{\phi}

\def\f{\phi}

\def\d{\delta}

\def\vr{ \textbf{r}}

\def\vv{\textbf{v}}
\def\va{\textbf{a}}
\def\oot{{1\over 2}}

\def\grad{\vec\nabla}

\def\gf{\grad\phi}

\def\c{\gamma}

\def\rar{\rightarrow}
\def\pd#1#2{{\partial{#1}\over\partial{#2}}}

\begin{document}

\TitreGlobal{Mass Profiles and Shapes of Cosmological Structures}

%%-----------------------------

%%      the top matter

%%-----------------------------

\title{MOND AS MODIFIED INERTIA}

\author{Milgrom, M.}\address{The Weizmann Institute Center for Astrophysics}

\runningtitle{}

\setcounter{page}{23}

% Keep this line, even if the page will be settled afterwards..

\index{Milgrom, M.}

% Repeat the authors here, this will help to make the final index

%\maketitle

%

\begin{abstract}

I briefly highlight the salient properties of modified-inertia
formulations of MOND, contrasting them with those of
modified-gravity formulations, which describe practically all
theories propounded to date. Future data (e.g. the establishment
of the Pioneer anomaly as a new physics phenomenon) may prefer one
of these broad classes of theories over the other. I also outline
some possible starting ideas for modified inertia.

 \end{abstract}
\maketitle

%

%%-----------------------------

%%      your text

%%-----------------------------

\section{Modified MOND inertia vs. modified MOND gravity}

 MOND is a modification of non-relativistic dynamics
involving an acceleration constant $\az$. In the formal limit
$\az\rar0$ standard Newtonian dynamics is restored. In the deep
MOND limit, $\az\rar \infty$, $\az$ and $G$ appear in the
combination $(G\az)$. Much of the NR phenomenology follows from
this simple prescription, including the asymptotic flatness of
rotation curves, the mass-velocity relations (baryonic
Tully-fisher and Faber Jackson relations), mass discrepancies in
LSB galaxies, etc.. There are many realizations (theories) that
embody the above dictates, relativistic and non-relativistic.
\par
The possibly very significant fact that $\az\sim cH_0\sim
c(\Lambda/3)^{1/2} $ may hint at the origin of MOND, and is most
probably telling us that a. MOND is an effective theory having to
do with how the universe at large shapes local dynamics, and b. in
a Lorentz universe (with $H_0=0,~ \Lambda=0$) $\az=0$ and standard
dynamics holds.
\par
We can broadly classify modified theories into two classes (with
the boundary not so sharply defined): In modified-gravity (MG)
formulations the field equation of the gravitational field
(potential, metric) is modified; the equations of motion of other
degrees of freedom (DoF) in the field are not. In modified-inertia
(MI) theories the opposite it true. More precisely, in theories
derived from an action modifying inertia is tantamount to
modifying the kinetic (free) actions of the non-gravitational
degrees of freedom. Local, relativistic theories in which the
kinetic actions are of the standard form with some physical metric
are of the MG type; so, relativistic MI theories are non-local or
non-metric.

\par
Start, for example, from the standard NR action
$$ S=-{1\over 8\pi G}\int\drt~(\gf)^2-\sum_i m_i\f(\vr_i)
+\sum_i m_i\int dt ~v_i^2(t)/2, $$ which describes a system of
masses $m_i$ interacting gravitationally. Modifying gravity would
be modifying the free action of the gravitational potential (the
first term) into something like $-(\az^2/8\pi
G)\int\drt~F(\az,\f,\gf, ...)$, where in the deep MOND limit
$F\propto \az^{-3}$ (e.g. the theory of Bekenstein and Milgrom
1984). In MI we replace the particle kinetic action by
 $~\sum m_i S_K[\az,\{\vr_i(t)\}]$, where $\{\vr_i(t)\}$ represents
 the full trajectory of particle $i$ and the kinetic action is a
 functional of it. In the deep MOND limit
  $S_K\rar{1\over \az}s_K[\{\vr(t)\}].$
In such theories the equation of motion of a particle in the
(unmodified) gravitational potential, $\f$, is of the form
  $~ \textbf{A}[\{\vr(t)\},\vr(t),\az]=-\gf[\vr(t)],$
  where the inertia functional $ \textbf{A},$ of the dimensions of acceleration,
   is a functional of the whole trajectory
  and a function of the instantaneous position; it reduces to the
   acceleration for $\az\rar 0$. For
  $\az\rar\infty$ the equation of motion takes the form
 $\textbf{U} [\{\vr(t)\},\vr(t)]=-\az\gf[\vr(t)].$

 Special relativity entails a familiar example of modified (non-MOND)
 inertia with the standard NR particle kinetic action being replaced by
  $S_K=-\int\d\tau=-\int[1-(v/c)^2]^{1/2}~dt$
such that the equation of motion becomes
 $\textbf{F}=md(\c\vv)/dt=m\textbf{A}=m\c[\va+\c^2(\va\cdot\vv)\vv/c^2].$

\par
With the exception of some heuristic proposals described in
Milgrom (1994, 1999), all MOND theories proposed to date are of
the MG type (e.g. Bekenstein \& Milgrom 1984, Soussa \& Woodard
2003, Bekenstein 2004, Sanders 2005).

\section{Some properties of non-relativistic modified inertia theories}
In Milgrom (1994, 1999) I derived certain general properties of NR
MI formulations of MOND for particle dynamics: If we retain
Galilei invariance in addition to the requirements of Newtonian
and MOND limits, the particle kinetic action has to be non-local
in time. For example, an action of the form $\int f(a/\az)v^2~dt$
can give the desired MOND dynamics, but is not Galilei invariant.
The Lorentz invariant action $-\int F(a^\mu a_\mu/\az^2)d\tau$
($a^\mu=d^2x^\mu/d\tau^2$), replacing the Lorentz free particle
action $-\int d\tau$, does have a Galilei invariant NR limit, but
this is, alas, $-\int F(a^2/\az^2)dt$, which is not the correct NR
action. It seems to me that if we forgo Galilei invariance we
should replace it with a more general symmetry, one that involves
$\az$, and that reduces to Galilei when $\az\rar 0$. This must
then entail a corresponding extension of Lorentz invariance (see
below).

 Given a particle kinetic action, $S_K$, bound trajectories
satisfy an integral, virial relation of the form
$S_K(1+\pd{lnS_K}{ln\az})=\oot\langle\vr\cdot\gf\rangle$
($\langle\rangle$ is the time average). From this follows that for
any circular orbit in an axi-symmetric potential we have

$$\mu(g/\az)g=g_N,$$ where $g=v^2/r$ is the correct (MOND) acceleration,
$G_N=-\partial \f/\partial r$ the Newtonian acceleration, and
$\mu(x)$ is simply derived from  the action as restricted to
circular orbits (we only have to know the action values for
circular orbits to get $\mu(x)$).

%  $S_K= {1\over 2T}\int_{-T}^T\vv\cdot
%f[{S_K\overleftarrow{D}_t\overrightarrow{D}_t\over \az^2}]\vv~dt$
% EOM: $(1+\oot\pd{lnS_K}{\az})f(-{S_KD^2_t\over
%\az^2})D^2_t\vr(t)=-\gf$

\section{Observable differences}
While the most salient aspects of galaxy dynamics are very similar
in mondified inertia and mondified gravity, there are important
difference that may eventually help reject one in favor of the
other.

1. The predictions of the two differ when forces other than
gravity are present; e.g., in a Milliken-like experiment where
strong gravity is almost balanced by an electric force, resulting
in a sub-$\az$ acceleration. In MG there should not be a MOND
departure, as the gravitational field is large; in MI there
should, as the total acceleration is small. Such an experiment
does not seem feasible at present.

2. The definition of conserved quantities, and  adiabatic
invariants, (momentum, angular momentum, etc.) in terms of the
non-gravitational degrees of freedom is different in the two
approaches: these quantities are derived from the kinetic actions,
which are modified in MI, but not in MG (for example, in SR the
momentum is $m\c\vv$). All significant tests of MOND to date
concern stationary situations and do not involve the conservation
laws. But future studies involving formation, mergers, accretion,
relaxation, etc. of and in galaxies may become accurate enough to
constrain the type of underlying modification.

3. Even in simple stationary situations, predictions of
observables, such as galaxy rotation curves, may differ somewhat
in the two classes of theories. For example, we saw above that MI
predicts $\mu(g/\az)g=g_N$ for the rotation curves, while MG (e.g.
the NR modified-Poisson theory propound by Bekenstein and Milgrom
1984) give somewhat different results. The differences were
considered by Brada \& Milgrom (1995); they are not large and are
also partly masked by uncertainties in the form of the
interpolating function $\mu$. But, with the number of galaxies
with good data increasing, time may be ripe for a detailed
analysis that might constrain $\mu(x)$ and simultaneously perhaps
distinguish between the alternatives (see e.g.  Famaey \& Binney
2005).

4. With MG we still have in the NR regime $\va\equiv\dot\vv=-\gf;$
so all test bodies have the same acceleration at the same position
in the modified potential $\f$ irrespective of their trajectory.
With MI, the inertial force per unit mass $\textbf{A}$ is not the
acceleration
 anymore; so the measured acceleration depends not only on
position but on details of the trajectory as well. (In SR, e.g.,
electrons running perpendicular or parallel to an electric field
have the same $d(\c\vv)/dt$, but undergo different accelerations.)
In particular, the function $\mu(a/\az)$ appearing above in the
description of circular orbits in MI is not relevant for other
trajectories, for which we do not even have a simple relation
between the MOND and Newtonian accelerations. For instance, a term
in the action of the form $\int~dt~f(a/\az)(\va\cdot\vv/\az)^2$
enters strongly for linear trajectories, but  does not affect
circular trajectories at all (since it vanishes for them). The
fact that in MI we have to specify an action that is a functional
of the trajectory permits us an infinitely larger freedom then in
MG. So we can make the modification strongly dependent on orbital
eccentricity, or on the degree of binding of the orbit, etc. etc..
\par
In galaxies, one measures instantaneous velocities and distances,
assumes an orbit, and deduces the acceleration from these. If it
were possible to directly measure the accelerations of bodies in
the same position but on different orbits they should agree in MG
but may differ in MI. It is difficult to estimate the expected
differences without a specific theory. In the Newtonian regime the
differences are small, of course, whereas in the NR MOND regime we
saw that the equation of motion is of the form $\textbf{U}
[\{\vr(t)\},\vr(t)]=-\az\gf[\vr(t)],$ where $\textbf{U}$ has
dimension of acceleration$^2$, and has the same value for all
particles at the same position. The differences in the actual
accelerations might then not be so strongly dependent on the orbit
if, for example, $\textbf{U}$ is dominated by $a^2$. Perhaps a
comparison between the behavior of massive bodies and light rays
will enlighten us on this point, but for that we would need a
relativistic version of MI.
\par
Closer to home, the Pioneer anomaly, if verified as a new-physics
effect (Anderson et al. 2002), might provide a decisive test. It
can be naturally explained in the context of MOND as MI but is
difficult to explain in the context of a MG theory (Milgrom 2002):
The Pioneer anomaly has no match in planetary motions for which a
constant, unmodelled acceleration of the magnitude shown by the
spacecraft is ruled out by a large margin. The planets probe
heliocentric radii smaller than where the Pioneer anomaly has been
found. So a MG theory may still have a little leeway by having the
anomaly set in rather abruptly with distance just at the interim
heliocentric radii (e.g., Sanders 2005). A MI explanation will
build on the fact that the orbits of the spacecraft differ greatly
from those of the planets: the former are close to linear and
unbound, the latter quasi circular and bound. It is intriguing in
this connection that the analysis for Pioneer 11 (Anderson et al.
2002) shows an onset of the anomaly just around the time where the
spacecraft was kicked from a bound, nearly elliptical orbit to the
unbound, almost linear orbit on which it is now (the corresponding
event for Pioneer 10 is not covered). The onset still wants
verification, but if real, it would be a signature of MI.
\par
In the sense discussed here, the dark matter doctrine is a kind of
MG; so any indication that the mass discrepancy in galactic
systems is due to MI will also argue against DM.

\section{Possible approaches to MOND inertia}
We do not have a MI theory for MOND at the level of satisfaction
achieved for for MG formulations. This line of inquiry has
attracted relatively little attention, perhaps because MI is
technically more difficult to implement as a fundamental theory.
But, instances of MI in effective theories are rife in physics,
from the kinematics of electrons in solids and bodies in fluids,
to mass renormalization and the Higgs mechanism in field theory.
MOND too could result as such an effective theory. Special
Relativity is another possible source of inspiration in seeking to
modify inertia. It entails a modification of newtonian inertia,
brought about by the imposition of a new symmetry: Lorentz
invariance. Whichever idea we follow we should be guided by the
cosmological connection of $\az$, hinting that MOND might result
only in the context of a non-Minkowskian universe, with $\az$
reflecting the departure from flatness of space time.

\subsection{derived, effective inertia}

It is well known that objects moving in a medium with which they
interact (electrons in solids, photons in refractive media, bodies
in fluids) may exhibit a revised form of inertia. Surprisingly, it
often happens that the interactions with the medium can all be
encapsuled, at some level of approximation, as a reshaping of the
inertial properties of the object: its motion is governed by a
modified, effective ``free'' action with the degrees of freedom of
the medium disappearing from the problem. MOND inertia, or indeed
the whole of inertia, may result in a similar way. We then have to
find an appropriate omnipresent medium, describe the interaction
of all known physical DoF with it, and show that to a sufficient
approximation this interaction can lead to inertia as we know it
(with MOND). In other words, we want to show that the known
actions of all DoF result as effective actions from such a
mechanism. This would shed new light on Mach's principle because
MOND brings into account a new connection between the universe at
large and inertia.
\par
An effective theory can violate some of the hallowed principles of
relativity even though the fundamental theory from which it is
derive does not: Effective theories may be non-local, violate the
equivalence principle at different levels, etc.. An effective
theory also has a more limited applicability than its parent
theory. So, if we derive an effective MOND inertia as we now apply
it to galactic systems, with the acceleration constant and the
interpolating function coming out of the model in the context of
cosmology, this theory need then not be applicable to cosmology
itself (perhaps not even to local systems involving strong
gravitational fields). The hope is, however, that when we
understand the origin of MOND in such terms, the role played by
the inertia-modifying medium and the way it affects cosmology and
other strong-field systems can also be understood.
\par
As discussed in Milgrom (1999), the vacuum might constitute an
appropriate medium: we know it can define an inertial frame since
an accelerated observed can detect it's acceleration with respect
to the vacuum through the Unruh effect. And the vacuum is also
affected by the cosmological state of the universe (e.g., the
Gibbons Hawking effect) so it has the potential to explain the
nonzero $\az$ as a result of non-Minkowskian cosmology. (The field
or medium responsible for the observed acceleration of the
universe is also a potential candidate: its deduced present day
density is numerically related to $\az$, which could underlie the
link.) I presented in Milgrom (1999) a heuristic argument showing
how a MOND-like inertia could follow in this context. There are
also pieces of evidence suggesting that kinetic actions can form
spontaneously solely through interaction of DoF with the vacuum.
For example, the mere interaction of the electromagnetic field
with the charged DoF of the vacuum produces a contribution (of the
standard form) to its kinetic action--the so called
Heisenberg-Euler action (see e.g. Itzykson and Zuber 1980). But we
are still a far cry from having a theory based on this idea.
\par
Some general questions arise when one embarks on such a program:
The known instances of derived inertia start from standard
physics; so all degrees of freedom start with their standard
inertia, which is then modified by the interaction with the
medium. Is MOND then also a correction on a preexisting inertia?
Are there two contributions to inertia, one the standard, whose
origin is just assumed by fiat, and another that modifies it into
the MOND form? Or is there only one origin to inertia giving the
standard form at $\az\rar 0$ and MOND at the other end? I suspect
the latter because in the formal limit $\az\rar \infty$ inertia
disappears; so; it may require fine tuning to have the two
contributions to inertia cancel in the limit, standard inertia
being independent of $\az$. (But the MOND correction could also be
multiplicative, in which case this argument is neutralized.)
\par
And, if inertia is to be produced totally from scratch, does that
include the purported inertia-endowing medium itself? In the
instances we have of derived inertia, the Newtonian inertial law
is still obeyed exactly, and the difference between the effective
inertial force and the actual rate of change of momentum of the
object is taken up by the medium. This means that the medium
itself can have momentum, hence must have inertia to begin with.
It remains to be seen whether real-world inertia can be produced
with a medium itself devoid of it.
\par
Another course of research in this vein is to construct mechanical
models for inertia based on well understood physics, such as the
inertia that is acquired by bodies moving in fluids; then to see
in this framework whether MOND-like behavior can result in a
context resembling cosmology. If successful this will tell us at
least that the above program is feasible, and will perhaps teach
us how to go about it.

 \subsection{New symmetries}
In another approach we may try to construct MOND inertia on lines
similar to those of special relativistic inertia, which follows
from Lorentz invariance of the kinetic action. We could then seek
a new symmetry that forces a form of the free actions compatible
with MOND. (See, for example, an attempt by Kowalski-Glikman \&
Smolin 2004 along such lines, using an extension of SR having two
more constants beside the speed of light--so called ``triply
special relativity''.)
\par
 What is the space on which this new
symmetry acts? Is it still space-time or a larger one?  The
extended, or modified, symmetry should appear, according to the
cosmological connection of MOND, because we live in a
cosmologically curved space-time; it should then disappear or
return to Lorentz invariance when $\az\rar 0$. Presumably $\az$ is
to play the role similar to that of of the speed of light in SR
whose appearance as a limiting speed has to do with the
Minkowskian signature of space-time. But, in contrast, $\az$ is
not a limiting acceleration and there are no discontinuities as we
cross it. This may be telling us that we should be looking for
rotations between axes that span a manifold with Riemannian
signature.
\par
Without having a concrete application in mind, I am personally
intrigued by the following observations, which may give some
reader a clue in the right direction. A de Sitter Universe (dSU),
which approximates our universe as it is at present, is a
maximally symmetric space time with positive curvature and
Minkowskian signature. It can be viewed as a 4-D pseudo-sphere
embedded in a flat 5-D Minkowski space, $M^5,$ centered at the
origin, say. Consider an arbitrary, time-like world line
$x^\mu(\tau)$ in the dSU having a local acceleration $a^\mu\equiv
D^2 x^\mu/D\tau^2$, of magnitude $a=(-a^\mu g_{\mu\nu}
a^\nu)^{1/2}$. Then the acceleration in the $M^5$ embedding space
$a_5^A\equiv d^2x^A/d\tau^2$ has magnitude $a_5=(-a_5^A \eta_{AB}
a_5^B)^{1/2}$, which can be shown to be related to $a$ by
$a_5=(a^2+c^2\Lambda/3)^{1/2}$. Above, $g_{\mu\nu}$ is the metric
in the dSU, $\eta_{AB}$ that of $M^5$, and $\Lambda=3/R^2$ the
cosmological constant specifying the curvature radius, $R$, of the
dSU. So, if we make the connection with MOND by defining
$\haz=c(\Lambda/3)^{1/2}$ to play a similar role to $\az$, we can
write $a_5=(a^2+\haz^2)^{1/2}$.
\par
Inertial world lines, with $a^\mu=0$, are time-like geodesics of
the dSU: great pseudo circles, which are the intersects of the dSU
with (2-D) planes though the origin in the embedding space. It can
be shown that world lines of finite, constant acceleration $a$ are
the intersects of the dSU with planes at a (Minkowskian) distance
$d$ from the origin with
$d/R=a/(a^2+\haz^2)^{1/2}\equiv\l(a/\haz)$. For a body at some
point $p$ on its world line compare two observers whose world
lines go through $p$ and are tangent there to the body's world
line, one is inertial, the other has the same acceleration, $a$,
as our body at $p$. These two reference world lines correspond to
two planes one through the origin and one a distance $R\l(a/\haz)$
from the origin ($p$ itself is by definition a distance $R$ from
the origin). We can transform one plane to the other by a rotation
through $p$ by an angle $\theta$ with $sin\theta=\l(a/\haz)$. So,
kinematic factors such as $\l(a/\haz)$ resembling MOND's
$\mu(a/\haz)$ appear in this contexts as geometrical quantities:
matrix elements of a rotation taking one from an inertial observer
to an accelerated one, just as the Lorentz factor $\c$ appears in
the context of Lorentz transformations. Perhaps, in a similar
manner, such factors can find their way into the equation of
motion of particles to give a desired MOND behavior.

\end{document}